\documentclass[preprint,showpacs,preprintnumbers,amsmath,amssymb]{revtex4}
\usepackage{graphicx,epsfig}
\usepackage{dcolumn}
\usepackage{bm}
\begin{document}
\title{
Are there the
VP couplings in the $\psi(3770)$ non-charmed decays hidden
behind the current measurements?}
\author{D. Zhang, G. Rong, J.C. Chen \\
  \em Institute of High Energy Physics, Beijing 100049, China}
\email{zhangdh@mail.ihep.ac.cn}

\begin{abstract}
  {A global analysis of the full amplitudes for $e^+e^-
  \rightarrow$ VP (Vector and Pseudoscalar) channels at $\sqrt{s}$
  =3.773 GeV and 3.670 GeV, which were measured by the CLEO-c
  Collaboration, shows that those measurements are essentially
  nontrivial for searching for the $\psi(3770)$ non-$D\overline D$
  decays. Unlike the nearly negative verdict on the $\psi(3770)$
  strong decays to the VP channels in the original analysis of the
  CLEO-c data, there exist some unusual solutions that predict
  the remarkable strength of $SU(3)$ symmetry VP decay of $\psi(3770)$
  resonance, which give some clue to understand the mechanism of
  $\psi(3770)$ non-$D\overline D$ decays and to reexplain the well-known
  $\rho-\pi$ puzzle in the J/$\psi$ and $\psi(3686)$ decays.}  
\end{abstract}
\pacs{13.20.Gd, 13.66.Bc, 14.40.Gx}
\maketitle

\section{Motivation}
There is a long-standing puzzle in understanding the exist measurements
for $\psi(3770)$ and $D \overline D$ production cross sections at the
peak of $\psi(3770)$ production in $e^+e^-$ annihilation \cite{nondd,R-Z-C}.
Potential Models predict that $\psi(3770)$ decays into
$D \overline D$ with branching fraction of $\sim 100\%$.
Recently careful investigation shows that the branching fraction of
$\psi(3770)$ non-$D\overline D$ decay would be up to more
than $10\%$ \cite{PhysRevLett_97_121801, PhysLettB_641_145}.
It is very interesting to know what are the exclusive
$\rm{non-}D \overline D$ final states of $\psi(3770)$ decays. 
Except the electromagnetic transitions and hadronic transitions
of $\psi(3770)$ to lower charmonium states,
are there indeed other significant exclusive
non-charmed decay modes from $\psi(3770)$ decays?

In the charmonium decays, there is another long-standing puzzle in understanding
the $\rho\pi$ decays of J/$\psi$ and $\psi(3686)$. The partial widths of
$\rho\pi$ channel and other VP channels in the
$\psi(3686)$ decays are unexpectedly lower than those in
J/$\psi$ decays. This is so called ``$\rho\pi$" puzzle. Are the
J/$\psi$ decay rates enhanced by some unknown mechanism or the
$\psi(3686)$ decay rates are suppressed abnormally? To investigate
the possible source of this puzzle, it is also important to measure
the $\psi(3770)$ VP decay amplitude.

Recently, the BES Collaboration \cite{BES_KK} observed a large
production cross section for $e^+e^- \rightarrow K^*(892)^0
  \overline K^0$+c.c.
\begin{equation}
\sigma(e^+e^-\rightarrow {K^{*0}\overline{K}^0}+c.c.)
  =(15.0 \pm 4.6\pm3.3)~~{\rm {pb}},
\nonumber
\end{equation}
at center-of-mass energy of $\sqrt{s}$=3.773 GeV and found that
the $K^{*\pm}(892) K^{\mp}$ production is suppressed.
Taking into account the possible interference between the
strong decay amplitude and the continuum production amplitude at
$\sqrt{s}$=3.773 GeV, the BES Collaboration set an upper limit
on the strong decay partial width for
$\psi(3770)\rightarrow K^*(892) \overline{K}$+c.c. to be
\begin{equation}
 \Gamma(\psi(3770)\rightarrow K^*(892)
    \overline{K}~+~\rm {c.c.}) < 29.0 ~~{\rm keV}
\nonumber
\end{equation}
at $90\%$ confidence level.

The CLEO-c Collaboration made more careful studies of twelve
exclusive VP decay channels for $\psi(3770)\rightarrow \rho\pi$, 
$K^*(892) \overline{K} + c.c.$, $\omega\pi^0 $, $\rho\eta $,
$\rho\eta' $, $\omega\eta$, $\omega\eta'$, $\phi\eta$,
$\phi\eta' $ and $\phi\pi^0$ reported in Ref. \cite{CLEO_c_VP}. The
CLEO-c Collaboration measured the cross sections for all of the channels
at the energies $\sqrt{s}$=3.773 GeV and $\sqrt{s}$=3.670 GeV.
The CLEO-c results show that the measured cross sections at $\sqrt{s}$=3.773
GeV are almost equal to or even less than the ones measured at
$\sqrt{s}$=3.670 GeV, which mean that the net cross sections for the
$\psi(3770)$ decays are consistent with zero except
only for the channel $\psi(3770)\rightarrow \phi\eta$. 
The negative results about the $ \psi(3770)$ strong VP decays led people
ignoring the important strong decay component existing in $\psi(3770)$
and only focusing their attention on the form factors of those
channels as well as the isospin violation in electromagnetic
interaction \cite{CLEO_c_VP}.

In this paper, we develop a model to account for both the amplitudes of
electromagnetic (E-M) production and $\psi(3770)$ strong decay in the
process of $e^+e^-\rightarrow$ VP. By analyzing the cross sections for
the exclusive VP channels, which were measured by the CLEO-c Collaboration,
we extract out the branching fractions for $\psi(3770)$ decay to these VP
final states.

\section{The Model and the Formulae}
In the $\psi(3686)$ decay sector, because of the smallness of the strong
VP decay coupling, the E-M decay component as well as the continuum (E-M)
component of the VP channel would be no longer the small amounts
comparing with those of strong decay. People have to deal with the two
components properly \cite{Wang_2003}. 
At the resonance peak, the production amplitude consists of two
parts, one is the decay amplitude of charmonium resonance and the
other is continuum E-M production amplitude. In the
resonance decay part, there are two components as well. They are the E-M decay
amplitudes and the strong decay amplitude.
Totally, there are three components involved in the $e^+e^-$ annihilation
process at the resonance peak, which are the strong decay component, the E-M decay
component and the continuum production component.

Unlike the VP decays of J/$\psi$ and $\psi(3686)$, the E-M decay amplitudes
of $\psi(3770)$ can be neglected due to the little tiny dileptonic
decay branching fraction. There are only the strong decay amplitude and the
continuum production amplitude in the $e^+e^-$ collision production at
$\sqrt{s}$=3.773 GeV. Typically, according to the conventional
point of view, the partial widths of the $\psi(3770)$ VP decay
channels could be up to keV order of magnitude, like their cousins in J/$\psi$
decays. However, due to the large width of $\psi(3770)$ the decay amplitudes
of those channels can not get large amplification as the ones at the narrow
resonance states. Associated with the measurements of the form factors of channel
$\omega\pi^0$ at the energies of $\psi(3686)$ and $\psi(3770)$ resonance
vicinities \cite{BES_ompi0,CLEO_3686_VP,CLEO_c_VP} the decay process with
only a few keV partial width of the
rather wide resonance is really hard to be measured if one does not consider
the interference between the amplitudes of the strong decay and the
continuum production. As the measurements by CLEO-c \cite{CLEO_c_VP},
both the evident yield excess of channel $\phi\eta$ and the rather large
yield deficit of the $\rho\pi$ channel at the resonance peak
hint that there must be rather complex interference between the two kinds
of amplitudes acting globally on the VP channels. Some destructive
interference just shows up at $\rho\pi$ channel in the ``deficit" way.
And more complex interferences cause the $\phi\eta$ yields enhanced at
resonance peak. In fact, in such complicate interference case the decay
contributions may easily be covered up by the continuum contribution and
the interference contributions. If one completely neglects the buried decay
contribution, the single E-M amplitude assumption would not describe the
measured cross sections well. In practice, it is dangerous to measure the
branching fractions for the $\psi(3770)$ non-$D\overline D$ decays by simply
considering the net yields for the channels observed at the peak of $\psi(3770)$
over that at the nearby off resonance region.
In this analysis we introduce the strong decay amplitudes in the analysis
formalism to see how the strong decay affects the VP production at the
$\psi(3770)$ resonance peak.

We describe the global decay of $\psi(3770)$ and the continuum production
process still based on the flavor $SU(3)$ invariant model, which was
developed thirty years ago \cite{haber_su3}, In this model
the strange quark mass correction in both of the strong coupling and
the E-M coupling, the wave function nonet symmetry breaking and the double
Okubo-Zweig-Iizuka (DOZI) suppression effects are all taken into account.
As for the continuum production at the two energy points $\sqrt{s}$=3.773 GeV and 3.670 GeV,
except the coherent strong decay amplitudes from the J/$\psi$ and $\psi(3686)$
tails, which can safely be neglected from the calculations, there is only the
continuum E-M amplitudes itself.
In addition, we guess that the incoherent
component contributions which are mainly from the initial state radiative
(ISR)  return to J/$\psi$ and
$\psi(3686)$ resonances have efficiently been rejected in the work reported
in Ref. \cite{CLEO_c_VP} and can be neglected in our analysis too. 

Following the convention given in Ref. \cite{Seiden_1988},
we define that ${\bf g}$ represents the VP strong decay amplitude
in the flavor $SU(3)$ symmetry limit; 
${{\bf {g}}_s}$ represents the strong decay amplitude from $s$ quark,
\begin{equation}
s_g=1-({\bf {g}}_s/|{\bf {g}}|)/2,
\label{s_g}
\end{equation}
characterizes the $SU(3)$ mass violation, which is as the same as the
parameter ``$s$" given in the Tab. VIII of Ref. \cite{Seiden_1988};
$\theta_{P}$ represents the $\eta-\eta'$ mixing angle; the product
$r\cdot {\bf {g}}$ represents the amplitude correction of the $SU(3)$ nonet symmetry
violation with the factor $(1-s_{\rm P})$ for a strange pseudoscalar production
and with the factor $(1-s_{\rm V})$ for a strange vector production. If $s_{\rm V}
=s_{\rm P}$=0, (exactly $s_{\rm V}+s_{\rm P}$=0), $r\cdot {\bf {g}}$
measures the pure DOZI amplitude correction.
Unlike the case in Ref. \cite{Seiden_1988}, because the E-M amplitude is no
longer small comparing with the strong amplitude, we have to consider both
the isoscalar and isovector components of the E-M amplitude. We define
the E-M amplitude in the form of $SU(3)$ octet matrix representation as  
\begin{equation}
 {\bf E}={\bf e}_1\cdot \rm {I}_3+{\bf e}_0\cdot \rm Y  
\nonumber
\end{equation}
in which ${\bf e}_0$ and ${\bf e}_1$ are the isoscalar and isovector
components, respectively,
and $\rm {I}_3$ and Y are, respectively, the isospin third component and the
hypercharge matrices in flavor $SU(3)$ octet space. We define $\theta_0$
as the phase of ${\bf e}_0$ relative to ${\bf {g}}$, $\delta_1$ as the
phase shift difference of ${\bf e}_1$ to ${\bf e}_0$ and a factor $(1/2-s_e)$
as the correction for strange quark coupling to E-M isoscalar part ${\bf {e}}_0$.
We assume that the couplings ${\bf e}$'s and their phases do not change
in the all VP channels, and their moduli at the two different energy points
$\sqrt{s}$=3.773 GeV and $\sqrt{s}$=3.671 GeV only change with a $1/s^3$
dependence. If ${\bf e}_1={\bf e}_0$, we return to common definition as
Refs. \cite{haber_su3,Seiden_1988} did.

\begin{table*}[htbp]
\caption{The Amplitudes for VP production in $e^+e^-$ annihilation
at $\sqrt{s}=3.773$ GeV. The coupling
${\bf{g}}_K$ defined in channels $K^{*}\bar K+c.c.$ can be considered
as a free parameter if one of ${\bf{g}}$ and ${\bf{g}}_s$ is fixed.}
\begin{center}
\begin{tabular}{c|cc}  \hline
Channel($ch$)                        &$M^{ch}_{res,3770}$                                                                     &$M^{ch}_{ctm,3770}$                  \\ \hline
$\rho^0\pi^0,\rho^{\pm}\pi^{\mp}$    &${\bf {g}}$                                                                             &${\bf {e}}_0$                        \\
$\omega\eta$                         &${\bf {g}}X_{\eta}+\sqrt{2}r{\bf{g}}[\sqrt{2}X_{\eta}+(1-s_{\rm P})Y_{\eta}]$           &${\bf {e}}_0X_{\eta}$                \\
$\phi\eta$                           &${\bf {g}}_sY_{\eta}+r{\bf {g}}(1-s_{\rm V})[\sqrt{2}X_{\eta}+(1-s_{\rm P})Y_{\eta}]$   &$-2{\bf {e}}_0(1-s_e)Y_{\eta}$       \\
$\omega\eta'$                        &${\bf {g}}X_{\eta'}+\sqrt{2}r{\bf{g}}[\sqrt{2}X_{\eta'}+(1-s_{\rm P})Y_{\eta'}]$        &${\bf {e}}_0X_{\eta'}$               \\
$\phi\eta'$                          &${\bf {g}}_sY_{\eta'}+r{\bf{g}}(1-s_{\rm V})[\sqrt{2}X_{\eta'}+(1-s_{\rm{P}})Y_{\eta'}]$&$-2{\bf {e}}_0(1-s_e)Y_{\eta'}$      \\
$\omega\pi^0$                        & 0                                                                                      &$3{\bf {e}}_1$                       \\
$\phi\pi^0$                          & 0                                                                                      &0                                    \\                                             
$\rho^0\eta$                         & 0                                                                                      &$3{\bf{e}}_1X_{\eta}$                \\
$\rho^0\eta'$                        & 0                                                                                      &$3{\bf{e}}_1Y_{\eta}$                \\
$K^{*0}\bar K^0+c.c.$                &${\bf {g}}_K=({\bf {g}}+{\bf {g}}_s)/2$                                                 &$-{\bf {e}}_0(1/2-s_e)-3/2{\bf e}_1$ \\
$K^{*\pm}K^{\mp}$                    &${\bf {g}}_K=({\bf {g}}+{\bf {g}}_s)/2$                                                 &$-{\bf {e}}_0(1/2-s_e)+3/2{\bf e}_1$ \\ \hline
\end{tabular}
\label{tab_amp}
\end{center}
\end{table*}

For the channels ``$ch$", ($ch$=$\rho^0\pi^0,K^{*0}\overline{K}^0+c.c.$, etc)
at energy $\sqrt{s}$,
$M^{ch}_{res,\sqrt{s}}$ denotes the resonance decay amplitude and
$M^{ch}_{ctm,\sqrt{s}}$ denotes the
continuum production amplitude.
The total production amplitudes are then written as
\begin{equation}
M^{ch}_{\sqrt{s}}=M^{ch}_{res,\sqrt{s}}+M^{ch}_{ctm,\sqrt{s}},
\label{tot_am}
\end{equation}
in which
\begin{equation}
M^{ch}_{res,3670}=0,
\nonumber
\end{equation}
and
\begin{equation}
M^{ch}_{ctm,3670}=M^{ch}_{ctm,3770}\cdot f_d
\nonumber
\end{equation}
at the energy of $\sqrt{s}$=3.670 GeV,
where $f_d=3.773^3/3.670^3$ is the scaling factor for the $1/s^3$
energy dependence of the cross section. The amplitudes
$M^{ch}_{ctm,3770}$ and $M^{ch}_{res,3770}$
defined at $\sqrt{s}$=3.773 GeV for all of the channels
are listed in Tab. \ref{tab_amp}.
In the table, $X_{\eta}={\rm{cos}}(54.736^o+\theta_{P})$,
$Y_{\eta}={\rm{sin}}(54.736^o+\theta_{P})$,
$X_{\eta'}=-{\rm{sin}}(54.736^o+\theta_{P})$ and
$Y_{\eta'}={\rm{cos}}(54.736^o+\theta_{P})$, which are the same
as those given in Ref. \cite{Seiden_1988}.
Those amplitudes completely control the correlations
among the VP channel productions. If any significant decay amplitude
$M^{ch}_{res,3773}$ given in Tab. \ref{tab_amp}
has been measured to be non-zero,
which means that the measured couplings ${\bf {g}}$, ${\bf {g}}_s$ etc. are non-zero,
this indicates that $\psi(3770)$ has a significant branching fraction 
for decay to the non-charmed channel ``$ch$".

\begin{table*}[htbp]
\caption{The numbers of the events observed by CLEO-c, the
subscripts $sw$ and $sb$ indicate the signal window and the side
band window, respectively; the upper script 3.67 and 3.77 indicate
the c.m. energies.}
\begin{center}
\begin{tabular}{c|cc|cc|c} \hline
Channel($ch$)         &$N^{3.67}_{sw}$&$N^{3.67}_{sb}$&$N^{3.77}_{sw}$&$N^{3.77}_{sb}$& $\epsilon (\%)$\\ \hline
$\rho\pi$             & 43           &    5.4       &314           & 44.8         & 26.3           \\
$~\rho^0\pi^0$        & 21           &    3.4       &130           & 33.0         & 32.5           \\
$~\rho^{\pm}\pi^{\mp}$& 22           &    2.0       &184           & 11.8         & 23.1           \\
$\omega\pi^0$         & 54           &    6.2       &696           & 39.2         & 19.0           \\
$\phi\pi^0$           & 1            &    1.6       &2             & 40           & 16.5           \\
$\rho^0\eta$          & 36           &    3.1       &508           & 31.0         & 19.6           \\
$\omega\eta$          & 4            &    0.0       &15            &  6.0         &  9.9           \\
$\phi\eta$            & 5            &    1.0       &132           & 15.9         & 11.0           \\
$\rho^0\eta'$         & 1            &    0.0       &27            &  0.9         &  2.9           \\
$\omega\eta'$         & 0            &    0.0       &2             &  0.0         &  1.5           \\
$\phi\eta'$           & 0            &    0.0       &9             &  2.0         &  1.2           \\
$K^{*0}\bar K^0+c.c.$ & 38           &    0.4       &501           & 18.1         &  8.8           \\
$K^{*\pm}K^{\mp}$     & 4            &    1.0       &36            & 32.4         & 16.0           \\ \hline
\end{tabular}
\label{tab_cleo_c}
\end{center}
\end{table*}

We start the analysis from the observed numbers
of the events for the VP channels, their errors
and their corresponding detection efficiencies 
at the energies of $\sqrt{s}$=3.773 GeV and 3.670 GeV, which were
published by CLEO-c Collaboration \cite{CLEO_c_VP}. The numbers and
errors of the events are obtained in both the signal windows and the side
bands. Tab. \ref{tab_cleo_c} shows those numbers and detection efficiencies
which are given in Ref. \cite{CLEO_c_VP}. Taking
the numbers of the events from Tab. \ref{tab_cleo_c}, we obtain the
numbers,
\begin{equation}
N^{obs,ch}_{\sqrt{s}}=(N^{\sqrt{s}}_{sw}-N^{\sqrt{s}}_{sb})_{ch},
\label{N_obs}
\end{equation}
of the observed events at $\sqrt{s}$ for the channel ``$ch$".
Usually the determinations of the detection efficiencies and the ISR
corrections are all energy dependent and relate to
the production line shapes for those channels. We guess that the
determinations of the detection efficiencies given in Ref. \cite{CLEO_c_VP}
were done under the assumption of that the continuum cross section line shape
is in the $1/s^3$ energy dependence and the energy cut is at $\sqrt{s'} \geq
J/\psi$ mass, where $\sqrt{s'}$ is the center of mass energy of the ISR return
system. It should be stressed that, for the ISR and FSR (Final State Radiative)
corrections in the continuum processes, our calculation gives
$\eta_{ctm}$=1.19 at $\sqrt{s}$=3773 GeV and $\eta_{ctm}$=1.11 at
$\sqrt{s}$=3670 GeV, while Ref. \cite{CLEO_c_VP} gives
$\eta_{ctm}$=1/1.20=0.833 at both of the two energy points.
As for the resonance decay, the ISR correction is quite different
from the one for the continuum process. In our calculation, the ISR
correction factor for the resonance is $\eta_{res}$=0.824 including the
FSR correction at $\sqrt{s}$ =3.773 GeV for all channels.
In our ISR correction calculations, the $v^3$ phase space dependences
have been taken into account, where
$v=\sqrt{[1-(m_V+m_P)^2/s][(1-(m_V-m_P)^2/s]}$ is the
velocity of the vector daughter in the CM decay system.
However, in the calculation of $\eta_{ctm}$'s, a mean production threshold
of the channels has been set to serve as the common threshold for all of
those channels. For this reason, the channel dependences of the corrections
$\eta_{ctm}$'s are ignored in this analysis.
For the calculation of the contribution of the interference terms
among the amplitudes describing different processes, we have to know
their own detection efficiencies and ISR corrections. In the
analysis we simply take the geometric average of the related
coefficients as the effective ones. 

Using the amplitudes of VP channels in Eq.(\ref{tot_am}) and
Tab. \ref{tab_amp}, the efficiencies
for those channels, the ISR and FSR correction as given above,
the $v^3$ phase space dependences
and the luminosities $L_{\sqrt{s}}$ accumulated at
the two collision energy points of $\sqrt{s}$ =3.773 GeV and 3.670 GeV,
we can calculate the expected numbers $N^{exp,ch}_{3670}$ and
$N^{exp,ch}_{3773}$ of the event yields for those channels
under the simplification assumption concerning the interference terms
mentioned above. For channel ``$ch$"
\begin{equation}
N^{exp,ch}_{3670}=L_{3670}\cdot v^3_{3670}\cdot \epsilon^{ch}_{ctm}
\cdot\eta_{ctm,3670}^{ch}\cdot |M^{ch}_{ctm,3670}|^2
\nonumber
\end{equation}
and
\begin{eqnarray}
&&N^{exp,ch}_{3773}=L_{3773}\cdot v^3_{3773}\times \nonumber \\
&&
\left |\sqrt{\epsilon^{ch}_{res}
\cdot\eta_{res}^{ch}}\cdot M^{ch}_{res,3773}+\sqrt{\epsilon^{ch}_{ctm}
\cdot\eta_{ctm}^{ch}}\cdot M^{ch}_{ctm,3773}
\right |^2.
\nonumber
\end{eqnarray}

Comparing the numbers $N^{obs,ch}_{\sqrt{s}}$ defined in Eq.(\ref{N_obs})
with the expected one $N^{exp,ch}_{\sqrt{s}}$, we get the equation set
\begin{equation}
N^{exp,ch}_{\sqrt{s}}=N^{obs,ch}_{\sqrt{s}}
\label{general_eq}
\end{equation}
in which $ch$ = $\rho\pi$, $K^*(892) \overline{K}$+c.c., $\omega\pi^0$,
$\rho\eta$, $\rho\eta'$, $\omega\eta$,$\omega\eta'$,$\phi\eta$,
$\phi\eta'$ and $\phi\pi^0$, $\sqrt{s}$=3.670 and 3.773 GeV. Because of
the zero observation and zero expectation for the
$\phi\pi^0$ channel, listed in Tab. \ref{tab_cleo_c} and Tab. \ref{tab_amp},
we can get rid of this channel in our analysis.
So we only focus our attention on the rest eleven channels.

In the maximum likelihood fit, leaving the parameters ${\bf g}$, ${\bf g}_s$,
${\bf e_0}$, ${\bf e_1}$, $r$, $s_e$, $s_{\rm V}$, $s_{\rm P}$, $\theta_{P}$
$\delta_1$ and    
${\rm{cos}}\theta_0$ free,
we can solve the Eq.(\ref{general_eq}) by  maximizing the probability function
\begin{equation}
      Prob= \prod_{ch,\sqrt{s}}^{2N_{ch}}
P_{ch}(N^{obs,ch}_{\sqrt{s}},N^{exp,ch}_{\sqrt{s}})
\label{prob}
\end{equation}
where $P_{ch}(N^{obs,ch}_{\sqrt{s}},N^{exp,ch}_{\sqrt{s}})$
is the probability of finding $N^{obs,ch}_{\sqrt{s}}$ events with the
assumed mean number $N^{exp,ch}_{\sqrt{s}}$ at the energy $\sqrt{s}$, 
and get the solution of Eq.(\ref{general_eq}) with most probable values
of the parameters
which control the VP channel production at $\sqrt{s}$=3.773 and 3.670 GeV.

From the solutions of ${\bf {g}}$, $r$, ${\bf {e}}_0$ and/or
${\bf {e}}_1$ etc, one can get the Born cross sections for the VP channel
production at $\sqrt{s}$=3.773 or 3.670 GeV. For example, the Born cross
section for decay channel $\psi(3770)\rightarrow$ ``$ch$" at $\sqrt{s}$
=3.773 GeV can be written as
\begin{equation}
\sigma^{ch}_{res}=v^3_{3773}\cdot |M^{ch}_{res,3773}|^2.
\label{sigma_res}
\end{equation}
The total Born cross section for $e^+e^-$ annihilation to the channel ``$ch$"
at $\sqrt{s}$=3.773 GeV is given by
\begin{equation}
\sigma^{ch}_T=v^3_{3773}\cdot |M^{ch}_{res,3773}+M^{ch}_{ctm,3773}|^2. 
\label{sigma_tot}
\end{equation}
While the Born cross section for the channel ``$ch$" in continuum
production at $\sqrt{s}$ is then given by
\begin{equation}
\sigma^{ch}_{ctm,\sqrt{s}}=v^3_{\sqrt{s}}\cdot |M^{ch}_{ctm,\sqrt{s}}|^2.
\label{sigma_em}
\end{equation}

\section{The results}
According to the different coupling configurations of the
amplitudes for the VP channels, shown in Eq.(\ref{tot_am}) and 
Tab. \ref{tab_amp}, the twelve VP channels can be divided into
three sub-sets. The first
one consists of the channels without E-M isovector components, such
as $\rho\pi$, $\omega\eta$, $\phi\eta$, $\omega\eta'$ and $\phi\eta'$.
The second one consists of the E-M isovector component only, which is the pure E-M
channels $\omega\pi^0$, $\rho\eta$, and $\rho\eta'$.
The third subset includes only the channels $K^*\overline K+c.c.$, for which
the amplitudes
involve all of the two E-M coupling parts and the strong coupling component
as well as their interferences in the VP production. Since there is no
common coupling parameter in the first two sub-sets despite of the
pseudoscalar mixing angle $\theta_P$, one can simply try to solve
Eq.(\ref{general_eq}) separately in the two sub-sets at first.

\subsection{The channels without isovector E-M amplitude}
We start the analysis from the first set in which the channels are without
the isovector E-M amplitude contribution.
From the numbers of events, $N^{obs,ch}_{\sqrt{s}}$,
observed at the two energy points of $\sqrt{s}$=3.671 GeV and 3.773 GeV for
the six channels $\rho^0\pi^0$, $\rho^{\pm}\pi^{\mp}$, $\omega\eta$, $\phi\eta$,
$\omega\eta'$ and $\phi\eta'$, we solve the Eq.(\ref{general_eq}).
Leaving all of the related parameters, such as $|{\bf g}|$,
${\bf g}_s$, $|{\bf e}_0|$, $cos\theta_0$, $s_e$, $\theta_P$, $r$, $s_V$
and $s_P$ free, the fitting yields\\
$|{\bf {g}}|=2.752\pm{0.291}$,\\
${\bf {g}}_s=1.279\pm{0.706}$,\\
$\rm{cos}\theta_0=-0.935\pm{0.141}$,\\
$|{\bf {e}}_0|=1.495\pm{0.126}$,\\
$s_e=0.264\pm{0.247}$,\\
$r=-0.236\pm{0.070}$,\\
$\theta_{\rm P}=(-17.83\pm{12.12})^o$,\\
$s_V=0.092\pm{0.941}$\\
and\\
$s_P=-0.091\pm{0.587}$.\\
Indeed, as hinted by the measurements from the CLEO-c \cite{CLEO_c_VP} and guessed
above, the solution of $\rm{cos}\theta_0=-0.935\pm{0.141}$ shows that
${\bf{g}}$ and ${\bf{e}}_0$ are almost opposite in the VP
production at $\psi(3770)$ resonance peak.
Associated with earlier measurements in the J/$\psi$
production and decays, in which the E-M decay amplitudes and the strong decay
amplitude are more likely with the phase difference of $\phi_{e,res}-\phi_{g}\sim
+90^o$ \cite{kopke,achasov}, and the phase difference between the
continuum amplitude and the one of resonance E-M decay at the resonance
peak is $\phi_{e,ctm}-\phi_{e,res}=+90^o$ too \cite{BES_95, Wang_03},
the measurement of the phase difference $\theta_0=\phi_{e,ctm}-\phi_{g}
=(160^{+20}_{-17})^o$ between ${\bf{e}}_0$ and ${\bf{g}}$ here is reasonable.
The $90^o$ phase
difference between the E-M decay amplitudes and the strong decay amplitude
shown in J/$\psi$ decays \cite{kopke,achasov} is mainly due to the exist
of an original $90^o$ phase difference from the short distance force
range in the on-shell three gluon annihilation of $1^{--}$ quarkonium
states. This argument should be kept in the $\psi(3770)$ decays. Out of the hard
quark-gluon interaction level the strong phase shift of long distance final
state interaction should essentially be small \cite{zheng}.
The resolutions of $s_V=0.092\pm{0.941}$ and $s_P=-0.091\pm{0.587}$, which
are consistent with zero, seem to mean
the small nonet symmetry breaking between the singlet and octet wave functions.
\begin{table*}[htbp]
\caption{The fitted parameters.}
\begin{center}
\begin{tabular}{c|c|ccccc}  \hline
                     && {\bf{Solution 1}}        & {\bf{Solution 2}} 	 &{\bf{Solution 3}}            & {\bf{Solution 4}}       \\
                     && the channels            & the channels            & all channels               & all channels             \\
                     && without isovector       & with   isovector        &  with                      &  without                 \\
                     && E-M component           & E-M component           & counter terms              & counter terms            \\ \hline    
$|{\bf g}|$          &&$2.64^{+0.41}_{-0.54}$   &2.64(fixed)              &$2.67^{+0.38}_{-0.32}$      &$2.43^{+0.31}_{-0.29}$    \\
${\bf g}_s$          &&$1.23^{+0.67}_{-0.74}$   &$-1.79^{+0.42}_{-0.36}$  &$1.18^{+0.71}_{-0.78}$      &$-1.23^{+0.52}_{-0.49}$   \\
$s_g$                &&$0.27^{+0.14}_{-0.13}$   &$0.84^{+0.17}_{-0.20}$   &$0.28\pm{0.16}$             &$0.75\pm{0.42}$           \\
${\bf g}_s^{\rm{add}}$&&                         &                         &$-1.51^{+0.43}_{-0.40}$     &0.0(fixed)               \\
$\rm{cos}\theta_0$   &&$-0.91^{+0.12}_{-0.13}$  &-0.91(fixed)             &$-0.91\pm{0.07}$            &$-0.86^{+0.06}_{-0.04}$   \\
$|{\bf e}_0|$        &&$1.49^{+0.12}_{-0.13}$   &1.49(fixed)              &$1.49^{+0.12}_{-0.13}$      &$1.44^{+0.13}_{-0.14}$    \\
$s_e$                &&$0.25^{+0.19}_{-0.22}$   &$-0.35\pm{0.11}$         &$0.20^{+0.20}_{-0.23}$      &$-0.29^{+0.13}_{-0.14}$   \\
$a^{\mu}$            &&                         &                         &$-0.55^{+2.38}_{-0.22}$     &0.0(fixed)                \\
$|{\bf e}_1|$        &&                         &$1.223^{+0.016}_{-0.018}$&$1.226^{+0.017}_{-0.019}$   &$1.22\pm{0.02}$           \\
$\delta_1$           &&                         &$(7.39^{+5.90}_{-7.69})^o$&$(6.70^{+6.59}_{-8.19})^o$ &$(7.62^{+8.88}_{-6.02})^o$ \\
$\theta_P$           &&$(-18.7^{+5.8}_{-4.7})^o$&$(-23.2\pm{2.2})^o$      &$(-22.5^{+1.9}_{-2.0})^o$   &$(-24.1^{+2.1}_{-2.3})^o$ \\
$r$                  &&$-0.24\pm{0.06}$         &                         &$-0.26\pm{0.06}$            &$-0.32^{+0.09}_{-0.08}$   \\
$s_V$                &&0.0(fixed)               &                         &0.0(fixed)                  &$-0.42^{+0.36}_{-0.59}$   \\
$s_P$                &&0.0(fixed)               &                         &0.0(fixed)                  &$0.45\pm{0.70}$           \\
$\chi^2/n_{dof}$     &&4.63/5=0.93              &3.98/5=0.80              &9.26/11=0.84                &13.2/11=1.20              \\ \hline
\end{tabular}
\label{tab_parameter}
\end{center}
\end{table*}
So we can assume that the nonet symmetry maintains for the wave functions,
which means that the nonet symmetry violation is only in the DOZI coupling.
If we fix $s_V$= $s_P$ =0 in the fit, we get
the parameters $|{\bf g}|$, ${\bf g}_s$, $|{\bf e}_0|$,
$cos\theta_0$, $s_e$, $\theta_P$ and $r$, which are listed in the second column
(``{\bf Solution 1}") of Tab. \ref{tab_parameter},
where $\theta_{\rm P}=(-18.73^{+5.8}_{-4.7})^o$ indicates 
\begin{equation}
X_{\eta}=Y_{\eta'}=0.809^{+0.045}_{-0.064}~\rm{and}~Y_{\eta}=-X_{\eta'}=-0.588^{+0.068}_{-0.079}.
\nonumber
\end{equation}
The fit gives $\chi^2/n_{dof}=4.63/5=0.93$, which is also listed in
Tab. \ref{tab_parameter}.
With the ${\bf {g}}$ and ${\bf {g}}_s$, we obtain the $SU(3)$ mass
correction
\begin{equation}
s_g=0.268^{+0.140}_{-0.127}.
\nonumber
\end{equation}
Except for the three coupling strengths, $|{\bf e}_0|$,
$|{\bf g}|$ and $|{\bf g}_s|$, which have their own dynamics
in the higher energy position, the other relative correction parameters, 
$s_g$, $s_e$, $r$ and the mixing angle $\theta_{\rm P}$ obtained from the fit
are all reasonable comparing with those obtained from J/$\psi$ decays measured
by Mark-III and DM2 \cite{MARK3_DM2}. However,
from Eqs.(\ref{tot_am}) and (\ref{sigma_res}), we find that the strong decay
coupling $|{\bf g}|$
gives quite large cross section, branching fraction and the
partial width for $\psi(3770)\rightarrow \rho\pi$ decay, which are
\begin{equation}
\left .
\begin{aligned}
&
\sigma^{\rho\pi}_{res}=(18.2^{+6.11}_{-6.70})~~{\rm{pb}}
\\&
B^{\rho\pi}=(1.83^{+0.62}_{-0.67})\times 10^{-3}
\\&
\Gamma^{\rho\pi}=49.7^{+16.9}_{-18.3} ~\rm {keV}.
\end{aligned}
\right \}
\label{sig_rhopi_iso0}
\end{equation}
The partial width is almost in two order of magnitude higher
than that of the conventional typical partial width of the
J/$\psi$ VP decays. The latter one is at the order of 1 keV.
If we assume that the fraction of the width of $\psi(3770)
\rightarrow \rho\pi$ to the width of $\psi(3770)\rightarrow light~
hadrons$ is roughly as the same as the one in the J/$\psi$ decays,
the huge partial width of $\psi(3770)\rightarrow\rho\pi$ would predict
that about $10\%$ of $\psi(3770)$ decays to non-$D\overline D$ final states.
This predicted branching fraction for $\psi(3770)\rightarrow {non-}D\overline
D$ is almost as the same as the one measured by BES-II Collaboration \cite{
PhysRevLett_97_121801,PhysLettB_641_145}.
The large cross section $\sigma^{\rho\pi}_{res}$ as shown in
Eq.(\ref{sig_rhopi_iso0})
is a factor of more than 3 of the total $\rho\pi$ production cross section,
\begin{equation}
\sigma^{\rho\pi}_T=5.35 ^{+1.45}_{-1.58} ~~{\rm{pb}},
\label{sig_rhopi_tot_3770}
\end{equation}
obtained from Eq.(\ref{sigma_tot}). This cross section is
consistent with the measurement reported in Ref. \cite{CLEO_c_VP},
(see Tab. \ref{tab_results}). In the Tab. \ref{tab_results} we list all
of the production cross sections and the branching fractions for
$\psi(3770)\rightarrow$ VP predicted in this work, and also listed the
production cross sections of the VP channels given in Ref. \cite{CLEO_c_VP}
as the comparison.
As for the $\rho\pi$ production cross section at
$\sqrt{s}=3.670$ GeV in this measurement, we get
\begin{equation}
\sigma^{\rho\pi}_{T,3670}=6.34 ^{+1.05}_{-1.03} ~~{\rm{pb}},
\label{sig_rhopi_tot_3670}
\end{equation}
which is indeed higher than the one given in
Eq.(\ref{sig_rhopi_tot_3770}) at resonance peak.
Because of the different determinations of the ISR corrections in Ref.
\cite{CLEO_c_VP} and in our work, the decrease of the Born cross section of
$\rho\pi$ channel at $\psi(3770)$ resonance peak is not so large as that
obtained by CLEO-c Collaboration \cite{CLEO_c_VP} (see also
Tab. \ref{tab_results}).
Owing to the large cancellation between the two amplitudes
${\bf g}$ and ${\bf e}_0$, the large cross
section $\sigma^{\rho\pi}_{res}$ for $\psi(3770)\rightarrow\rho\pi$
disappeared without the global amplitude analysis.

\subsection{The pure E-M channels $\omega\pi^0$, $\rho^0\eta$
and $\rho^0\eta'$ plus the channel $K^*\overline K$+c.c}
We consider together the last two sub-sets of the channels in which the
amplitudes of E-M production contain isovector component. The isoscalar E-M
component and the strong decay coupling only serve in the channel
$K^*\overline K$+c.c., and the strong coupling here is with a combined form
${\bf{g}}_K=({\bf g}+{\bf g}_s)/2$.
Inserting the ten numbers of the events observed
at the two energy points of 3.670 GeV and 3.773 GeV for the rest five
channels into Eq.(\ref{general_eq}), leaving ${\bf e}_1$, $\delta_1$,
$s_e$, $\theta_{\rm P}$ and ${\bf g}_K$ free (or instead of ${\bf g}_K$,
leaving ${\bf g}_s$ free but fixing ${\bf g}$ at some reasonable value
such as $|{\bf g}|$=2.635 obtained from last solution independently), and fixing
$|{\bf e}_0|$=1.494 and cos$\theta_0$=-0.907 obtained also from last
solution in assumption of that there is no
more correction added to the couplings ${\bf g}$ and $|{\bf e}_0|$ measured
in last solution in the channel $K^*\overline K$, we fit the numbers of events
observed in the five channels and obtain the solution of the free parameters.
The results are listed in the third column (``{\bf Solution 2}")
of Tab. \ref{tab_parameter}. The fit gives $\chi^2/n_{dof}=3.98/5=0.80$.

From above two solutions,
we see that the measured pseudoscalar mixing angles $\theta_{\rm P}$
in the two independent measurements are consistent with each other.
The isovector E-M component, ${\bf{e}}_1$, is really split from the
isoscalar one, ${\bf{e}}_0$, with almost 2$\sigma$ deviation in
magnitude and with a non-zero phase shift difference. However,
we note that the $s$ quark strong decay
coupling ${\bf {g}}_s$ and its E-M coupling correction $s_e$ in
the solution 2 are both with the negative values, while they are
supposed to be positive in the conventional
$SU$(3) invariant model with simple static mass corrections. The minus
$s_e$ is formally very likely to be some anomalous ``magnetic moments" term added to the
$s$ quark E-M coupling. As for the minus ${\bf g}_s$, which is obviously
irrelative to the three pure E-M channels, it seems that the $s$ quark
strong coupling undergo almost $180^o$ phase shift
from the conventional $SU(3)$ strong interaction wave function.
The odd behavior of the minus ${\bf g}_s$ as well as the E-M ``anomalous
magnetic moments" term for the $s$ quark indicate that there might be some
other dynamic sources or more complicated interaction correction contributing
to the production of $K^*\overline K$+c.c. in $e^+e^-$
annihilation.

Owing to the cancellation of ${\bf {g}}$ and the opposite ${\bf {g}}_s$
(small ${\bf g}_K$),
the decay cross section, branching fraction and partial width
\begin{equation}
\left .
\begin{aligned}
&
\sigma^{K^*\overline K~}_{res}=(0.558^{+0.702}_{-0.379})~~{\rm{pb}}
\\&
B^{K^*\overline K~}=(0.56^{+0.70}_{-0.38})\times 10^{-4}
\\&
\Gamma^{K^*\overline K~}=(1.52^{+1.92}_{-1.03}) ~~{\rm {keV}},
\end{aligned}
\right \}
\label{sig_ksk_res_iso1}
\end{equation}
for $\psi(3770)\rightarrow K^*\overline K$+c.c in this solution are quite small
comparing with that of $\psi(3770)\rightarrow\rho\pi$ measured in last
solution given in Eq.(\ref{sig_rhopi_iso0}). These results are also
listed in Tab. \ref{tab_results}.
The remarkable increase of the E-M coupling
$e_0$ by negative correction $s_e$ and the large interferences between
the two E-M amplitudes and the strong amplitudes result a serious
asymmetry between the total production cross sections of channels
$K^{*0}\overline K^0$+c.c. and $K^{*\pm} K^{\mp}$. For example,
at $\sqrt{s}=3.773$ GeV the cross sections
\begin{equation}
\left .
\begin{aligned}
&
\sigma^{K^{*0}\overline K^0~+c.c.}_T=19.30^{+3.55}_{-1.85}~~{\rm{pb}}
\\&
\sigma^{K^{*\pm}
K^{\mp}~}_T<0.50~~({\rm{pb,~~at~90\%~confidence~level}}),
\end{aligned}
\right \}
\label{sig_ksk_tot_iso1}
\end{equation}
are consistent with the observed values reported
in Refs. \cite{CLEO_c_VP} and \cite{CLEO_3686_VP}.
Using the parameters $|{\bf e}_1|$ and $\theta_{\rm P}$, and from
Eq.(\ref{sigma_em}) we get the production cross sections of the
three pure E-M channels,
\begin{equation}
\left .
\begin{aligned}
&
\sigma^{\omega\pi^0}_{ctm}=(11.74^{+0.31}_{-0.34})~~{\rm{pb}}
\\&
\sigma^{\rho\eta}_{ctm}=(8.00{\pm{0.22}})~~{\rm{pb}}
\\&
\sigma^{\rho\eta'}_{ctm}=(2.57{\pm{0.28}})~~{\rm{pb}}.
\end{aligned}
\right \}
\label{sig_ompi_etc_iso1}
\end{equation}
These are also listed in Tab. \ref{tab_results}.
The E-M production cross sections and the total production cross sections
obtained in this subsection are all systematically lower than those measured
by CLEO-c Collaboration \cite{CLEO_c_VP} by about $30\%$, (see Tab.
\ref{tab_results}). Those differences are also due to the different
determinations of ISR corrections in the two works as mentioned above.

\subsection{The global fit including all of the measured VP channels}
We can introduce two additional effective counter terms, ${\bf{g}}^{\rm{add}}_s$
and ${\bf {a}}^{\mu}$, to compensate the odd behavior appeared in the channel
$K^*(892) \overline{K}$+c.c. for both the strong coupling and E-M coupling
of $s$ quark. We can simply assume that ${\bf{g}}^{\rm{add}}_s$ and ${\bf
g}$, ${\bf {a}}^{\mu}$ and ${\bf e}_0$ are collinear, respectively.
In this case, we define the amplitudes of channels $K^*(892) \overline{K}$+c.c.
as
\begin{equation}
\begin{aligned}
&
M^{K^{*+} K^-,K^{*-} K^+}_{res,3770}=({\bf {g}}
+{\bf{g}}_s)/2+{\bf{g}}^{\rm{add}}_s,
\\&
M^{K^{*+} K^-,K^{*-} K^+}_{ctm,3770}=-{\bf
{e}}_0(1/2-s_e)+3/2{\bf e}_1+{\bf {a}}^{\mu},
\\&
M^{K^{*0}\overline{K}^0+c.c.}_{res,3770}=({\bf {g}}+{\bf {g}}_s)/2 
+{\bf g}^{\rm{add}}_s,
\\&
M^{K^{*0}\overline{K}^0+c.c.}_{ctm,3770}=-{\bf {e}}_0(1/2-s_e)
-3/2{\bf e}_1+{\bf {a}}^{\mu},
\end{aligned}
\label{am_add_ksk}
\end{equation}  
as given in Tab. \ref{tab_amp}. 
With those amplitudes we globally solve the Eq.(\ref{general_eq})
with all of the observed channels.
Fixing $s_V$= $s_P$ =0 and leaving all of the other parameters including
${\bf{g}}^{\rm{add}}_s$ and ${\bf {a}}^{\mu}$ free, the fit gives
the most probable values of parameters
$|{\bf {g}}|$, ${\bf {g}}_s$, ${\bf {g}}^{\rm{add}}_s$,
$\rm{cos}\theta_0$, $|{\bf {e}}_0|$, $s_e$, ${\bf {a}}^{\mu}$
$|{\bf {e}}_1|$, $\delta_1$, $r$, $\theta_{\rm P}$. Those results are
listed in the fourth column (``{\bf Solution 3}") of Tab. \ref{tab_parameter}
with $\chi^2/n_{dof}$=9.26/11=0.84. From the parameters of solution 3
and Eqs.(\ref{tot_am}),(\ref{sigma_res}),(\ref{sigma_tot}),
(\ref{sigma_em}) and (\ref{am_add_ksk}), we can get the production cross
sections of the twelve VP channels, including the zero measurement of
channel $\phi\pi^0$. For example, for the channels $\rho\pi$ and
$K^*(892)\overline{K}$+c.c., the decay cross sections from $\psi(3770)$
resonance, and their decay branching fractions and partial widths can
be calculated as
\begin{equation}
\left .
\begin{aligned}
&
\sigma^{\rho\pi}_{res}=(18.68^{+4.12}_{-4.19})~~{\rm{pb}}
\\&
B^{\rho\pi}=(1.87^{+0.41}_{-0.42})\times 10^{-3}
\\&
\Gamma^{\rho\pi}=51.1^{+11.2}_{-11.5} ~\rm {keV}
\\&
\sigma^{K^*\overline K~}_{res}=(0.56^{+1.28}_{-0.22})~~{\rm{pb}}
\\&
B^{K^*\overline K~}=(0.56^{+1.28}_{-0.22})\times 10^{-4}
\\&
\Gamma^{K^*\overline K~}=(1.52^{+3.50}_{-0.60}) ~~{\rm {keV}}.
\end{aligned}
\right \}
\label{sig_hshe_rhok}
\end{equation}
And the production cross sections of the three pure E-M channels
can be calculated as
\begin{equation}
\left .
\begin{aligned}
&
\sigma^{\omega\pi^0}_{ctm}=(11.79^{+0.34}_{-0.36})~~{\rm{pb}}
\\&
\sigma^{\rho\eta}_{ctm}=(7.92\pm{0.41})~~{\rm{pb}}
\\&
\sigma^{\rho\eta'}_{ctm}=(2.69\pm{{0.29}})~~{\rm{pb}}.
\end{aligned}
\right \}
\label{sig_hshe_ompi_etc}
\end{equation}
The measured values in Eqs.(\ref{sig_hshe_rhok}),(\ref{sig_hshe_ompi_etc})
are all consistent with those measured in the last two subsections.
As for the measurements of other channels, we have
\begin{equation}
\left .
\begin{aligned}
&
B^{\omega\eta}=(2.12^{+0.75}_{-0.48})\times 10^{-4}
\\&
B^{\omega\eta'}<0.25 \times 10^{-4},~({\rm~{at~90\%~c.l.}})
\\&
B^{\phi\eta}=(0.87^{+0.75}_{-0.58})\times 10^{-4}
\\&
B^{\phi\eta'}<0.60\times 10^{-4},~({\rm~{at~90\%~c.l.}}),
\end{aligned}
\right \}
\label{sig_hshe_phet_etc}
\end{equation}
obtained in the assumption of that there exist the counter terms ${\bf {g}}^{\rm{add}}_s$
and ${\bf {a}}^{\mu}$ and $s_V$=$s_P$=0.
If we fix the strong couplings ${\bf g}$ and ${\bf g}_s$ to be the
values which force the partial widths of $\rho\pi$ and $K^*\overline K$
etc. to be at the order of 1 keV which is at the same order of the J/$\psi$
VP decay coupling dynamics, the fitted $\chi^2$ is 47.27 for 15 degree of
freedom which corresponds to 5.3 standard deviation worse than that of
solution 3. 

If there essentially were no the two counter terms specially for channel
$K^*\overline K$+c.c (${\bf {g}}^{\rm{add}}_s={\bf {a}}^{\mu}=0$), i.e.
there existed the negative coupling ${\bf{g}}_s$ and negative correction
$s_e$ universally allowed on all of other relative channels with the value $s$
quark, the fit gives a somehow poor solution with large $\chi^2$=19.2
for 13 degree of freedom, which gives
2.7 standard deviation away from the counter term assumption (solution 3).
However, in this case, the effects of the opposite ${\bf {g}}_s$ and
negative $s_e$ in the related channels of ${\phi\eta}$ and ${\phi\eta}'$
might be compensated by somehow larger nonet symmetry breaking with non-zero
$s_V$ and $s_P$. Ignoring the terms
${\bf {g}}^{\rm{add}}_s$ and ${\bf {a}}^{\mu}$ used in the last solution and
leaving the nonet symmetry breaking parameters $s_V$ and $s_P$
free, we solve the Eq.(\ref{general_eq}) again to obtain a new solution.
The results are listed in the
fifth column (``{\bf Solution 4}") of Tab. \ref{tab_parameter}. The fit gives
$\chi^2/{\rm{n_{d.o.f.}}}$=13.2/11=1.2, which is a little bit higher than
those in other solutions. However, the solution still has the statistical
significance more than 5.6 standard deviation to the one obtained
in the assumption that only the E-M components (i.e. the ${\bf e}_0$, ${\bf e}_1$
and $\delta_1$, and the $\eta-\eta'$ mixing angle) act on the VP channel
production. As for the solution 4, we obtain the unusual solution again,
which is like solution 2 with the negative ${\bf {g}}_s$ and negative
E-M correction $s_e$ for the $s$ quark couplings. The parameters of
$r$, $s_{\rm V}$ and $s_{\rm P}$ which associate with the nonet symmetry
breaking measurements are now with unexpected larger values. Nevertheless,
as guessed above, those unusual numbers may suggest more complicate
dynamics for the $s$ quark production
either in the channel $K^*\overline K$ or in all of the relative channels,
involving value $s$ quark production.
This solution predicts the decay cross
sections, branching fractions and decay width of channels $\rho\pi$
and $K^*(892) \overline{K}$ and the production cross sections of three pure
E-M production channels $\omega\pi^0$, $\rho\eta$ and $\rho\eta'$,
\begin{equation}
\left .
\begin{aligned}
&
\sigma^{\rho\pi}_{res}=(15.50_{-3.52}^{+4.21})~~{\rm{pb}}
\\&
B^{\rho\pi}=(1.55^{+0.42}_{-0.35})\times 10^{-3}
\\&
\Gamma^{\rho\pi}=42.4^{+11.5}_{-9.6} ~\rm {keV}
\\&
\sigma^{K^*\overline K~}_{res}=(1.14^{+1.12}_{-0.57})~~{\rm{pb}}
\\&
B^{K^*\overline K~}=(1.14^{+1.12}_{-0.57})\times 10^{-4}
\\&
\Gamma^{K^*\overline K~}=(3.11^{+3.06}_{-1.56}) ~~{\rm {keV}}
\\&
\sigma^{\omega\pi^0}_{ctm}=(11.67^{+0.35}_{-0.38})~~{\rm{pb}}
\\&
\sigma^{\rho\eta}_{ctm}=(8.10^{+0.60}_{-0.61})~~{\rm{pb}}
\\&
\sigma^{\rho\eta'}_{ctm}=(2.43^{0.39}_{-0.37})~~{\rm{pb}},
\end{aligned}
\right \}
\label{sig_hs0he0_svsp_rhokompi}
\end{equation}
which are consistent with those obtained in the last solution with the counter
terms. However, for the channels $\omega\eta$, $\omega\eta'$, $\phi\eta$
and $\phi\eta'$, we obtain the branching fractions of
\begin{equation}
\left .
\begin{aligned}
&
B^{\omega\eta}=(0.92^{+0.75}_{-0.92})\times 10^{-4}
\\&
B^{\omega\eta'}<1.66 \times 10^{-4},~({\rm~{at~90\%~c.l.}})
\\&
B^{\phi\eta}<1.57\times 10^{-4},~({\rm~{at~90\%~c.l.}})
\\&
B^{\phi\eta'}=3.46^{+8.93}_{-1.75}\times 10^{-4},
\end{aligned}
\right \}
\label{sig_hs0he0_svsp_phet_etc}
\end{equation}
which are quite different comparing with those given in
Eq.(\ref{sig_hshe_phet_etc}).
As for the branching fractions for $\psi(3770)\rightarrow \phi\eta$,
the measured values obtained by this analysis
are also somehow inconsistent with that measured
by CLEO-c Collaboration \cite{CLEO_c_VP}, (see Tab. \ref{tab_results}).
However both the results show the large $\psi(3770)$ VP decay couplings.  
The summed total cross sections over all VP channels of $\sigma^{ch}_T$
at both energy points of $\sqrt s$=3.670 GeV and 3.773 GeV are listed
in Tab. \ref{tab_results}. The summed values of solution 3 and solution 4
as well as the results of CLEO-c measurement show in almost equals
at the two energy points. The real resonance decays have been
``hidden". This is mainly owing to the destructive nature of
interference between the couplings ${\bf g}$ and ${\bf e}_0$. 

\begin{table*}[htbp]
\tiny
\caption{Summary of the results. The errors for our measurements are
statistical only. The branching fraction of channel $\phi\eta$ measured
by CLEO-c Collaboration is directly from work \cite{CLEO_c_VP} and the
upper limits of other channels measured by CLEO-c Collaboration are
from the ``Method II" of paper \cite{CLEO_c_VP}}
\begin{center}
\begin{tabular}{c|cccccc}  \hline
                     &               & {\bf{Solution 1}}       & {\bf{Solution 2}}	 & {\bf{Solution 3}}          & {\bf{Solution 4}}       & {\bf{CLEO-c}}    \\ \hline
$\rho\pi$          &$\sigma_{3670}$ (pb)&$6.86^{+1.14}_{-1.12}$&                         &$6.89^{+1.61}_{-1.60}$      &$6.43^{+1.53}_{-1.35}$   &$8.0^{+1.7}_{-1.4}\pm0.9$ \\
                     &$\sigma_T$ (pb)&$5.35^{+1.45}_{-1.58}$   &                         &$5.42^{+3.38}_{-2.00}$      &$5.08^{+3.57}_{-1.26}$   &$4.4\pm0.3\pm0.5$ \\
                     &$\sigma_R$ (pb)&$18.24^{+6.11}_{-6.70}$  &                         &$18.68^{+4.56}_{-3.36}$     &$15.50^{+5.04}_{-2.96}$  &$<0.04$           \\
                     &Br $10^{-3}$   &$1.83^{+0.61}_{-0.67}$   &                         &$1.87^{+0.46}_{-0.34}$      &$1.55^{+0.50}_{-0.30}$   &$<0.004$          \\ \hline
$K^{*0}\bar K^0+c.c.$&$\sigma_T$ (pb)&                         &$19.30^{+3.55}_{-1.85}$  &$19.34^{+11.29}_{-6.78}$    &$19.44\pm3.61$           &$23.3\pm1.1\pm3.1$\\
                     &$\sigma_R$ (pb)&                         &$0.28^{+0.37}_{-0.19}$   &$0.28^{+0.64}_{-0.11}$      &$0.57^{+0.56}_{-0.29}$   &$<20.8$           \\ 
                     &Br $10^{-4}$   &                         &$0.28^{+0.37}_{-0.19}$   &$0.28^{+0.64}_{-0.11}$      &$0.57^{+0.56}_{-0.29}$   &$<20.8$           \\ \hline
$K^{*\pm}K^{\mp}$    &$\sigma_T$ (pb)&                         &$<0.502$                 &$<2.59$                     &$<0.77$                  &$<0.6$            \\
                     &$\sigma_R$ (pb)&                         &$0.28^{+0.37}_{-0.19}$   &$0.28^{+0.64}_{-0.11}$      &$0.57^{+0.56}_{-0.29}$   &$<0.1$            \\
                     &Br $10^{-4}$   &                         &$0.28^{+0.37}_{-0.19}$   &$0.28^{+0.64}_{-0.11}$      &$0.57^{+0.56}_{-0.29}$   &$<0.1$            \\ \hline
$\omega\pi^0$        &$\sigma_T$ (pb)&                         &$11.74^{+0.31}_{-0.34}$  &$11.79^{+0.33}_{-0.36}$     &$11.67^{+0.36}_{-0.38}$  &$14.6\pm0.6\pm1.5$\\
                     &$\sigma_R$ (pb)&                         &$0.0$                    &$0.0$                       &$0.0$                    &$<0.06$           \\
                     &Br $10^{-4}$   &                         &$0.0$                    &$0.0$                       &$0.0$                    &$<0.06$           \\ \hline
$\rho^0\eta$         &$\sigma_T$ (pb)&                         &$8.00\pm0.22$            &$7.92\pm0.41$               &$8.10\pm0.44$            &$10.3\pm0.5\pm1.0$\\
                     &$\sigma_R$ (pb)&                         &$0.0$                    &$0.0$                       &$0.0$                    &$<1.3$            \\
                     &Br $10^{-4}$   &                         &$0.0$                    &$0.0$                       &$0.0$                    &$<1.3$            \\ \hline
$\rho^0\eta'$        &$\sigma_T$ (pb)&                         &$2.57\pm0.28$            &$2.69\pm0.29$               &$2.43\pm0.32$            &$3.8^{+0.9}_{-0.8}\pm0.6$\\
                     &$\sigma_R$ (pb)&                         &$0.0$                    &$0.0$                       &$0.0$                    &$<0.4$            \\
                     &Br $10^{-4}$   &                         &$0.0$                    &$0.0$                       &$0.0$                    &$<0.4$            \\ \hline
$\omega\eta$         &$\sigma_T$ (pb)&$0.45^{+1.07}_{-0.23}$   &                         &$0.38^{+0.62}_{-0.20}$      &$0.32^{+0.42}_{-0.10}$   &$0.4\pm0.2\pm0.1$ \\
                     &$\sigma_R$ (pb)&$2.19^{+1.13}_{-0.74}$   &                         &$2.12\pm0.21$               &$0.92^{+0.92}_{-0.77}$   &$<0.1$            \\
                     &Br $10^{-4}$   &2.19                     & 			 &$2.12\pm0.21$               &$0.92^{+0.92}_{-0.77}$   &$<0.1$            \\ \hline
$\omega\eta'$        &$\sigma_T$ (pb)&$0.44^{+0.46}_{-0.26}$   &                         &$0.59^{+0.59}_{-0.39}$      &$0.44^{+1.84}_{-0.41}$   &$0.6^{+0.8}_{-0.3}\pm0.6$\\
                     &$\sigma_R$ (pb)&$<0.33$                  &                         &$<0.27$                     &$<0.47$                  &$<1.9$            \\
                     &Br $10^{-4}$   &$<0.33$                  &                         &$<0.27$                     &$<0.47$                  &$<1.9$            \\ \hline
$\phi\eta$           &$\sigma_T$ (pb)&$3.99^{+2.75}_{-1.27}$   &                         &$3.91^{+2.75}_{-1.27}$      &$3.70^{+2.78}_{-1.02}$   &$4.5\pm0.5\pm0.5$ \\
                     &$\sigma_R$ (pb)&$0.84^{+0.79}_{-0.58}$   &                         &$0.87^{+0.85}_{-0.52}$      &$<1.45$                  &$2.4\pm{0.6}$            \\
                     &Br $10^{-4}$   &$0.84^{+0.79}_{-0.58}$   &                         &$0.87^{+0.85}_{-0.52}$      &$<1.45$                  &$3.1\pm{0.7}$            \\ \hline
$\phi\eta'$          &$\sigma_T$ (pb)&$1.95^{+4.49}_{-2.83}$   &                         &$2.29^{+4.20}_{-1.76}$      &$1.69^{+1.37}_{-0.51}$   &$2.5^{+1.5}_{-1.1}\pm0.4$\\
                     &$\sigma_R$ (pb)&$<0.52$                  &                         &$<0.66$                     &$3.46^{+4.90}_{-1.76}$   &$<3.8$            \\
                     &Br $10^{-4}$   &$<0.52$                  &                         &$<0.66$                     &$3.46^{+4.90}_{-1.76}$   &$<3.8$            \\ \hline
All ``$ch$" summed   &$\sigma^{all}_{3670}$ (pb)&              &                         &$58.3\pm6.2$                &$62.3\pm6.9$             &$64.2^{+15.8}_{-6.8}$\\
                     &$\sigma^{all}_{T,3770}$ (pb)&            &                         &$54.4^{+13.0}_{-7.4}$       &$52.9^{+6.3}_{-4.1}$     &$64.6^{+4.5}_{-4.3}$ \\ \hline
\end{tabular}
\label{tab_results}
\end{center}
\end{table*}

\section{Discussion and Summary}
From above global analyses of the production cross sections
for twelve channels of $e^+e^-\rightarrow$ VP measured by
CLEO-c Collaboration at the $\psi(3770)$ resonance peak of $\sqrt{s}$
=3.773 GeV and at the energy of $\sqrt{s}$=3.670 GeV in the continuum
region, we obtained four different solutions of the parameters to control
the VP channel production.
From Tab. \ref{tab_results} we see that the measured branching fractions or
the production cross sections except for the channels involving the nonet
symmetry breaking are all consistent with each other in the four solutions.

It is remarkable that the large $SU(3)$ symmetry strong decay strength
$|{\bf g}|$ leads to the huge branching fraction of a level of $10^{-3}$
for the typical channel $\rho\pi$, which
corresponds to the decay width of two order of magnitude higher
than that in J/$\psi$ decays. The large strong decay coupling hidden
behind the VP channel production in the $e^+e^-$ annihilation at the $\psi(3770)$
resonance peak might help people to understand the sources of the
$\psi(3770)$ non-$D\overline {D}$ decays and give people some useful
information to reexplain the long-standing $\rho\pi$ puzzle in the $1^{--}$
charmonium state VP decays. Furthermore, the large strong decay coupling
with the opposite $s$-quark strong coupling and the minus E-M
correction $s_e$ or equivalently, the $s$-quark
``anomalous magnetic moments" ${\bf{a}}^{\mu}$,
required by the $K^*\overline K$ production as presented in those solutions with
different treatments and different assumptions in this work are all unusual
comparing with the conventional hard gluon annihilation
picture plus single $\psi(3770)$ resonance assumption.
If those measurements and analyses are all correct, one has to re-understand the
strong interaction dynamics which leads to the large Okubo-Zweig-Iizuka
rule violation in the vector meson $\psi(3770)$ decays and the
strange behavior of the $s$ quark couplings to the light hadron production
in the energy region around $\psi(3770)$ resonance.
It seems that people has to seriously consider the role of the long distance
strong interaction corrections including the
$D(D_s)$ meson exchange scheme to describe those anomalous phenomena
and the large Okubo-Zweig-Iizuka breaking in this energy region, like
many authors did \cite{Lipkin,Q.Zhao}.

The destructive interferences mechanism might manifest the possibility of
the significant existences of buried $\psi(3770)$ non-$D\overline D$ decays.
However, how do people understand the large {\bf net inclusive} non-
$D\overline D$ hadron branching fraction of $\psi(3770)$ decays measured
by BES Collaboration \cite{PhysRevLett_97_121801,PhysLettB_641_145} recently.
Phenomenologically, for example, if the behavior of the reversed $s$ quark
strong coupling appeared in VP channel is still maintained in those cases,
it would lead to constructive interference with the parallel
continuum E-M amplitudes and cause abundant strangeness meson production
at the $\psi(3770)$ peak, resulting in the large net cross section excess.
The one of the exceptions is the channel $K^*\overline K$+c.c., in which the
E-M production is due to the magnetic moments coupling. The minus E-M
coupling correction for $s$ quark or the counter term
${\bf{a}}^{\mu}$ as a special ``anomalous magnetic moments" enhances the E-M
production at the continuum region and leads to the observation of the equal
cross sections of channel $K^{*}\overline K$+c.c. at the two energy points
$\sqrt s$=3.670 and 3.773 GeV in work \cite{CLEO_c_VP}. This argument
can be cleared up by coming more precisely experimental measurements. 
Of course, another probable outlet would be that there are more complicated
structures or contents in the $\psi(3770)$ resonance scope which are evident
in a measurement of cross sections for $e^+e^-\rightarrow$ hadrons by the BES
Collaboration \cite{BES_two,BES_two_DD}. This measurement of the cross section
indicates that there are somehow complicate ``diresonance" structure instead
of the conventional single $\psi(3770)$ resonance assumption \cite{BES_two,
voloshin}. The exist of the extra substances might respond to the unusual
behavior of $\psi(3770)$ VP decays and the measured large non-$D\overline D$
branching fraction of $\psi(3770)$ decays \cite{PhysRevLett_97_121801,
PhysLettB_641_145}. 
\section{Acknowledgments}
This work is supported in part by the National Natural Science Foundation
of China under contract No. 10935007.

\end{document}